\documentclass[journal]{IEEEtran}
\usepackage{balance}
\IEEEoverridecommandlockouts

\usepackage{amsmath,amssymb,amsfonts}
\usepackage{cleveref}
\usepackage{algorithmic}
\usepackage{graphicx}
\usepackage{textcomp}
\usepackage{xcolor}
\usepackage{changepage}
\usepackage{url}
\usepackage[numbers,sort&compress]{natbib}

\usepackage{atbegshi}
\usepackage{adjustbox}

\def\BibTeX{{\rm B\kern-.05em{\sc i\kern-.025em b}\kern-.08em
    T\kern-.1667em\lower.7ex\hbox{E}\kern-.125emX}}

\crefformat{equation}{equation~(#2#1#3)}
\Crefformat{equation}{Equation~(#2#1#3)}

\begin{document}

\title{Game-Theoretic Framework for Private Data Sharing in Vehicular Networks}

\author{
    Yousef AlSaqabi,~\IEEEmembership{Member,~IEEE,}
    Yinan Zhou,
    Faisal Nawab,
    and Bhaskar Krishnamachari,~\IEEEmembership{Fellow,~IEEE}
}

\maketitle

\begin{abstract}  

We present a novel game-theoretic framework designed to enhance privacy and scalability in decentralized vehicular data collection systems. The proposed hybrid architecture comprises vehicles that supply sensor data, independent servers that process data via secure multiparty computation, a coordinator node that manages data flow, and data consumers that set economic incentives. Crucially, our framework ensures that only the data consumer can access the fully aggregated data, preventing individual raw data exposure and significantly reducing privacy risks. By integrating principles of the Stackelberg competition from game theory, our approach dynamically balances privacy and economic incentives, enabling vehicles to make participation decisions based on perceived privacy risks and incentives. We empirically validate our framework using real-world vehicular location data, quantifying privacy risks by evaluating the accuracy with which a potential adversary can reconstruct a vehicle's path using only a subset of the shared data. This paper details the development and deployment of a data-trading platform within this framework, introducing a practical and privacy-preserving marketplace for profitable vehicle data sharing. Through experiments and simulations, we evaluate the effectiveness of the system in preserving privacy and explore the dynamics that influence vehicle participation. Our findings highlight the robustness of the proposed framework in preserving privacy while supporting an active data market.

\end{abstract}


\section{Introduction}
The emergence of smart cities and the progression towards autonomous transportation systems underscore the growing importance of vehicular ad hoc networks (VANETs) and the Internet of Vehicles (IoV)~\cite{IoV}. Vehicles continue to serve as a method of transportation, but they are evolving beyond this basic function to become key components of a broader interconnected network; they generate, collect, store, process, and transmit massive amounts of data used to make driving safer and more convenient~\cite{Hao}. 

Autonomous vehicles and entities, such as traffic management systems or infotainment service providers, make use of this data to improve their services and operations. For example, autonomous vehicles use data not only from their own sensors, but also from other vehicles, allowing better decision-making that leads to safer and more efficient operations~\cite{ameen2020review}. Traffic management systems can use real-time data to detect congestion, improve route planning, and reduce travel times~\cite{Soto2021ASO}. Furthermore, infotainment service providers can provide personalized content tailored to user preferences and location data~\cite{cellular_v2x}. Looking ahead, private data sharing in VANETs is expected to offer a growing number of applications. However, this shift in the transportation landscape also brings with it a variety of complex challenges that require careful examination and innovative solutions.

The centralized nature of traditional vehicular network architectures poses some challenges. Centralized servers manage the majority of traffic loads, and their failure can lead to significant disruptions, making the system vulnerable~\cite{security_challenges_iov}. Embracing decentralization in network management and distributed computing solutions can offer numerous benefits, potentially resolving the issues associated with centralization by enhancing the resilience, scalability, and efficiency of the network~\cite{jiang_blockchain-based_2019}.

Simultaneously, the increasing network complexity and the sensitive nature of vehicular data raise critical questions about privacy and security~\citep{security_challenges_iov, Ogonji, chen2022, alsaqabi-hci}. The widespread sharing of vehicular data, such as vehicle locations and sensor readings, could lead to serious privacy violations and potential misuse~\cite{chen2022, garg2020survey}. Despite the growing body of research focused on secure communication and privacy protection in vehicular networks~\cite{security_challenges_iov, garg2020survey, bc_iov_survey, Blockchain_solutions_IoV, blockchain_security_services}, there are critical gaps to address these concerns while simultaneously incentivizing users to participate in data sharing. 

This paper discusses the need for and benefits of moving towards a decentralized data collection process to enhance privacy by proposing a novel game-theoretic network framework. Our approach decentralizes the data collection process by introducing distributed servers across the network. Vehicles would periodically send their data to one of these servers at random. Data consumers can then request specific information from the network. Upon receiving such a request, a coordinator identifies and queries the relevant servers. This initiates an aggregation process, where the servers employ secure multiparty computation (SMPC)~\cite{SMPC} to produce aggregated data containing the complete set of location and time-stamped data points. The aggregated data are then forwarded to the data consumer, fulfilling their request while maintaining privacy safeguards.

The application of this approach involves an economic model inspired by the Stackelberg competition~\cite{gametheory}, a concept in game theory, where the data consumer and the vehicle owners are the key players. In this game, the data consumer assumes the role of the leader, making the first move by defining the number of servers in the network, the compensation provided to the participating vehicles, and the frequency with which these vehicles share their data. The vehicle owners follow by evaluating the terms set by the data consumer and decide whether or not to participate in the data provision based on their privacy concerns and potential monetary gains from data sharing. 

We develop a data-trading platform that operates within the proposed framework, allowing vehicles to share driving data with consumers for profit while ensuring a level of privacy. This platform provides a practical setting to assess the efficiency and reliability of the system. Additionally, through simulations, we evaluate the system's ability to preserve privacy and investigate the factors influencing vehicle participation in this system.

\vspace{2mm}
The key contributions of this paper are as follows:
\begin{itemize}
    \item We present a hybrid decentralized network framework designed to improve data sharing privacy.
    \item We propose a novel game-theoretic economic model to optimize data sharing and monetary incentives between data consumers and vehicle owners.
\end{itemize}
\vspace{2mm}

The remainder of this paper is organized as follows: Section~\ref{relatedwork} provides a review of the related work. Section~\ref{model} outlines our system model.  Section~\ref{econ} details the economic models and equations used. Section~\ref{results} presents our experimental setup and results. Finally, Section~\ref{conclusion} summarizes the paper and suggests directions for future research.\textbf{}

\section{Related Work} \label{relatedwork}
This paper is an extension of our previous short conference paper~\cite{alsaqabi}. Although it retains the original motivation and problem formulation, it significantly expands upon the initial work by including an extensive results section that empirically tests and validates the assumptions and claims made in the earlier paper. This section covers three key topics in vehicular networks: private data sharing, game theory, and the integration of secure multiparty computation (SMPC) with blockchain.

\subsection{Private Data Sharing in Vehicular Networks}

The concept of data sharing in vehicular networks has been explored in various studies, each proposing unique solutions to address the challenges of privacy, security, and efficiency.

Several studies have proposed different methods to improve data privacy and security in vehicular networks. Kaiser et al. proposed an Open Vehicle Data Platform that uses Blockchain technology to ensure the privacy of vehicle owners and drivers during the exchange of vehicle and driving data~\cite{Kaiser}. Kong et al. proposed an efficient and location privacy-preserving sensory data sharing scheme with collision resistance in IoV~\cite{kong_privacy-preserving_2019}. Their scheme uses the modified Paillier Cryptosystem to achieve location privacy-preserving multidimensional sensory data aggregation. Fan et al. proposed a cloud-based mutual authentication protocol to ensure efficient privacy preservation in the IoV system~\cite{fan_cloud-based_2021}.

Some studies have focused on improving the efficiency of data sharing in vehicular networks. Lu et al.~\cite{lu_blockchain_2020} proposed a hybrid blockchain to improve the efficiency of data sharing in IoV. They proposed an asynchronous federated learning scheme for edge-level learning models and further improved the efficiency of federated learning by selecting participating nodes to minimize the total cost.

Some studies have proposed incentive mechanisms to encourage data sharing. Zhang and Xu proposed a mechanism that combines certificateless message authentication and blockchain incentives to provide anonymity and non-repudiation for traffic-related message reporters~\cite{zhang_blockchain-based_2022}. Yeh et al. proposed a fair and privacy-preserving data query system for vehicle networks based on blockchain technology~\cite{yeh_blockchain-based_2022}. Their scheme provides features of sustainable data accessibility and a large amount of data storage, as well as an incentive mechanism to encourage users to share their traffic information. 

Beyond technical privacy solutions, there is also extensive economic literature exploring privacy valuation and data-sharing incentives. Acquisti et al.~\cite{acquisti_economics} present a comprehensive review of the economic foundations underlying privacy valuations, emphasizing how users weigh the economic benefits of data sharing against privacy risks. Similarly, Ghosh et al.~\cite{ghosh2011selling} discuss incentive-driven economic frameworks that model privacy as a commodity, particularly in auction-based settings, highlighting the complexities involved in balancing individual privacy preferences and economic incentives.

Our study builds upon and extends these previous works by introducing a framework that not only offers robust privacy safeguards through decentralized data collection and SMPC, but also explicitly incorporates principles from the privacy economics literature. This dual technological and economic approach provides a direct monetary incentive system coupled with rigorous privacy preservation, encouraging active market participation and balancing the privacy interests of vehicle owners with the economic interests of data consumers.

\subsection{Game Theory in Vehicular Networks} 

Game theory has found niche applications within the field of vehicular networks that cater to specific use cases and scenarios~\cite{game_theory_vanet}. One such application is in the field of data security; Gupta et al. introduced a blockchain-secured authentication mechanism based on game theory for the Internet of Vehicles~\cite{gupta_game_2022}.

Another application of game theory in vehicular networks is in the area of computation offloading. Liwang et al. developed an opportunistic Vehicle-to-Vehicle computation offloading scheme based on game theory~\cite{liwang_game_2019}. They modeled the computation offloading scheme and pricing strategy as a Stackelberg game, taking into account various factors such as vehicular mobility models, V2V contact durations, computational capabilities, channel conditions, and service costs. Similarly, Hassija et al. used a game-theoretic approach to map service providers and consumers to perform offloading services in a cost-optimal way in V2V communication~\cite{hassija_dagiov_2020}.

Game theory has also been used to optimize resource allocation in edge computing. Xu et al. proposed a task offloading scheme based on fuzzy neural networks and game theory to minimize the latency of task processing of users in the case of limited edge server resources~\cite{Xu}. Other researchers have applied game theoretical approaches to solve the problem of estimating the edge of the user with respect to the app vendors~\cite{He} and to optimize the energy consumed and the total task executed time of the mobile user in mobile edge computing~\cite{Chen}.

The effectiveness of game-theoretic models in vehicular networks often relies on fundamental economic concepts. Samuelson and Nordhaus~\cite{econ_book} provide a comprehensive foundation on economic principles relevant to game theory, including diminishing returns, rational decision making, strategic interactions, market structures, and incentive design, all crucial for accurately modeling behaviors and decisions within vehicular networks. Additionally, specific utility functions commonly employed in game-theoretic network models, such as exponential saturation functions, have been effectively demonstrated in telecommunications and network economics literature. Shakkottai and Srikant~\cite{tele_econ}, for example, explore economic pricing strategies within networked environments, illustrating how similar utility functions accurately represent real-world consumer behavior and network interactions.

In the field of vehicular networks, the application of game theory has focused primarily on data security, computation offloading, and resource allocation. However, our research diverges from these established paths by leveraging the Stackelberg game theory to incentivize private data sharing. Although previous studies, such as that of Liwang et al.~\cite{liwang_game_2019}, have also utilized Stackelberg games, our work uniquely applies them to balancing data sharing incentives with privacy preservation. By explicitly incorporating foundational economic concepts and established network utility functions, our framework introduces a novel combination of monetary incentives and privacy-aware mechanisms, distinguishing it clearly from the existing literature.

\subsection{SMPC and Blockchain in Vehicular Networks}

Secure Multi-Party Computation (SMPC) and blockchain represent powerful complementary cryptographic technologies widely used to address security and privacy challenges within distributed systems~\cite{smpc_blockchain}, including vehicular networks. These technologies share core advantages, including decentralization, enhanced privacy, and trustworthiness, significantly reducing the risks associated with adversarial attacks and unauthorized data reconstruction~\cite{homomorphic, bc_iov_survey, alladi2022comprehensive}. Both technologies facilitate privacy-preserving collaboration and transparency in decentralized settings, making them particularly relevant in applications such as vehicular data aggregation and secure transactions.

Secure Multi-Party Computation (SMPC) is a cryptographic protocol that enables distributed parties to jointly compute functions without revealing their private inputs~\cite{SMPC}. It has been widely used in various security applications, including privacy-preserving machine learning~\cite{bost2014machine}, private set operations~\cite{Freedman}, and secure genomic computations~\cite{genome}. SMPC typically leverages cryptographic schemes such as homomorphic encryption, which allows computations directly on encrypted data without compromising confidentiality~\cite{homomorphic}.

SMPC has been used successfully in vehicular networks to address privacy and security challenges. Song et al.~\cite{SMPC_Vanet} proposed an SMPC-based anonymous authentication scheme to solve security and privacy problems for VANETs. The authors of~\cite{AutoMPC} proposed a cooperative control strategy involving SMPC that performs computations while increasing resilience towards latency and adversaries. A notable contribution is provided by Cramer et al.~\cite{LSS}, who demonstrated how verifiable secret sharing (VSS) and SMPC could efficiently utilize Linear Secret Sharing Schemes (LSSS), overcoming previous performance limitations and improving protocol efficiency and flexibility in practical cryptographic settings.

In parallel with SMPC, blockchain technology has also emerged as a pivotal tool to improve privacy and security in vehicular networks. Its several advantages, such as decentralization, constant availability, and anonymity, offer a significant enhancement to these networks~\cite{bc_iov_survey}. Using blockchain technology, the vehicular network ecosystem can be transformed into an environment that is reliable, secure, and maintains the privacy of its users. 

The authors of~\cite{alladi2022comprehensive} present a comprehensive survey on blockchain applications to protect vehicular networks, including data sharing, transportation, and authentication. Li et al.~\cite{blockchain_meets_vanet} designed a decentralized architecture using blockchain to address distrust, identity protection, and location privacy in VANETs. Amoretti et al.~\cite{PoL} presented a decentralized, infrastructure-independent proof of location scheme based on blockchain technology that guarantees location trustworthiness and the preservation of user privacy. Hu et al.~\cite{bc_datasharing} proposed a blockchain-based data sharing scheme that can guarantee location data sharing with user authorization while ensuring efficient and secure data transmission. Javed et al.~\cite{Javed2020BlockchainBasedSD} proposed a blockchain-based, resource efficient, and secure data sharing mechanism for VANETs, utilizing edge service providers and incentive mechanisms to encourage accurate and timely service. 

The integration of blockchain and SMPC addresses key limitations of isolated deployments, where blockchain enhances transparent and decentralized data management, and SMPC secures computation processes without compromising data privacy. Yang et al.~\cite{block_smpc} proposed a blockchain-based secure multi-party computation architecture for data sharing, leveraging blockchain’s decentralization, verifiability, and reliability to improve SMPC's efficiency. Their approach integrates an aggregator consortium for data storage, verification, and joint computation, further reinforcing data security and access flexibility. Raja et al.~\cite{AI_Blockchain_IoV} illustrated how blockchain smart contracts, informed by SMPC principles, could effectively optimize transaction verification and security in vehicular networks, significantly reducing trust issues and energy demands inherent in these environments. Another relevant work introduced by Gupta et al.~\cite{HEncryption} combined SMPC with homomorphic encryption to facilitate location privacy, allowing users to verify their presence without revealing their precise locations. 

Our research extends previous work by uniquely utilizing SMPC, specifically based on linear secret-sharing schemes, for securely aggregating comprehensive vehicle datasets across decentralized servers. Unlike earlier approaches that separately addressed isolated privacy or security aspects, our framework explicitly employs SMPC for comprehensive, privacy-preserving data aggregation. Additionally, we assume blockchain-based smart contracts to manage transparent, automated, and trustworthy interactions among vehicles, servers, and data consumers. While detailed implementation specifics of smart contracts are outside the scope of this paper, their assumed integration emphasizes our framework's commitment to transparent incentivization and security, significantly enhancing privacy, scalability, and robustness in vehicular network ecosystems.

\vspace{1mm}

\section{System Model} \label{model}
Our network framework integrates a hybrid architecture designed to optimize both privacy and operational efficiency. It consists of four key components: the vehicles that provide the data, the independent servers processing the data, a coordinator responsible for orchestrating aggregation rounds, and the data consumer collecting the processed data. 

\begin{figure}[!t]
    \centering
    \includegraphics[scale=0.44]{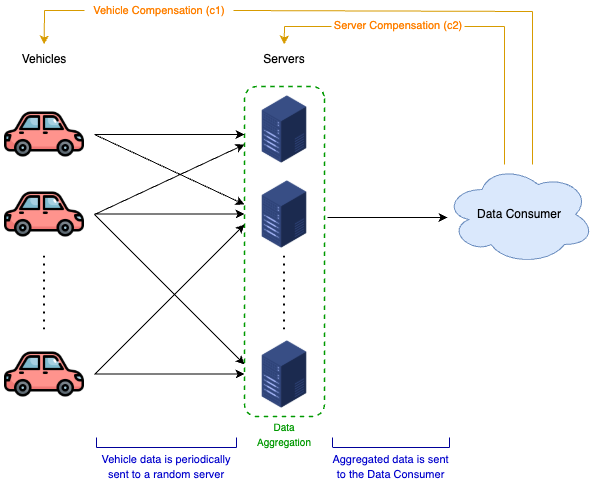}
    \caption{System Model}
    \label{fig:sys_model}
\end{figure}

Vehicles act as the main data providers in our network. As depicted in Fig.~\ref{fig:sys_model}, they periodically transmit their sensor data, which may include location coordinates or specific measurement readings, to one randomly selected server at a time. This approach ensures that no single server observes a vehicle's complete trajectory over time, thereby reducing the risk of path reconstruction from any individual server’s perspective.

Multiple data consumers can be active within the network, each with their own set of network parameters. These parameters influence which vehicles agree to share their data with a particular consumer, as described in Section~\ref{vehicles}. This selective sharing mechanism allows vehicles to participate in data transactions that align with their individual privacy preferences. 

When a data consumer requests specific information, it triggers a series of coordinated actions within the network. A coordinator oversees this process, managing the overall data flow without directly accessing or storing raw data. The coordinator represents a logical role rather than a single trusted physical server; in practical deployments, this coordination functionality can be replicated across multiple nodes using standard fault-tolerant consensus mechanisms to ensure availability and avoid reliance on a single point of failure.

The independent servers perform the core data processing tasks using Secure Multi-Party Computation (SMPC). SMPC enables multiple servers to jointly compute aggregated results from private inputs without revealing the raw data to other participants or external entities~\cite {SMPC}. Our framework utilizes SMPC based on linear secret-sharing schemes~\cite{LSS}: vehicle sensor data transmitted to servers is partitioned into multiple cryptographically secure shares, each of which individually reveals no meaningful information about the original data. During aggregation, servers collaboratively compute the desired aggregate function without reconstructing individual vehicle data. As a result, the privacy of raw sensor data is strictly maintained throughout the process. While alternative SMPC implementations, such as homomorphic encryption schemes~\cite{homomorphic}, could similarly ensure privacy, we chose linear secret-sharing schemes due to their computational efficiency, scalability, and reduced overhead in real-time vehicle network scenarios.

During initial testing and simulation phases, system-level logging (e.g., timing, message counts, and protocol state) may be enabled for debugging purposes. However, in operational deployment, the coordinator does not access raw vehicle trajectories. All aggregation is performed under the SMPC protocol, and the coordinator only receives aggregated outputs reconstructed from distributed shares.

We assume the use of blockchain-based smart contracts to foster autonomous, transparent, and trustworthy agreements among the three primary entities: vehicles, servers, and data consumers. These contracts automate the negotiation and agreement phases by recording commitments transparently, enforcing pre-agreed terms autonomously, and handling secure transactions efficiently. These smart contracts encourage broader participation from vehicle owners by providing transparent payment schemes and robust safeguards against fraudulent or inaccurate data submissions~\cite{smartcontracts, blockchain_security_services}. While blockchain ensures that interactions and data exchanges are verifiable and trusted, the detailed implementation of smart contract logic is beyond the scope of this work and left for future study. Transaction overhead can be amortized by batching settlements over longer intervals rather than issuing on-chain transactions per individual data sample.

\subsection{Vehicles as Privacy-Aware Data Providers} \label{vehicles}

Vehicles share private signed information, such as location data, with a random receiving server. In exchange, vehicles are compensated based on the amount of data they transmit. However, the higher the volume of data transmitted, the greater the privacy risk, given that the information sent includes time-stamped location data.

We model the utility of the vehicle by:

\begin{equation} \label{vutil}
    c_1\cdot f_d - L(f_d,s)
    \end{equation}

where $c_1$ is the payment provided to the vehicle, $f_d$ is the frequency of data transmission, and $s$ is the number of servers. We assume that all vehicles share data at the same frequency $f_d$. $L(f_d,s)$ is a function that says that privacy loss increases with $f_d$ and decreases with $s$. This loss function will be explained in more detail in Section~\ref{4b}.

We justify the use of our utility function in Eq.~\ref{vutil} as an economically intuitive representation, reflecting the proportional incentives vehicles receive based on their data contributions, balanced against a proportionally increasing privacy loss due to more frequent data exposure. Such incentive-driven utility functions are common in game-theoretic analyses of cooperative behaviors in vehicular networks, as highlighted by Sun et al.'s recent survey~\cite{game_theory_vanet}. 

Because of this trade-off, not every vehicle would be willing to contribute its data to the network; the utility gained would have to exceed the perceived privacy loss of sharing its data for the vehicle to have positive utility and, therefore, agree to contribute its data.  

This will create a supply curve because as $c_1$ increases, more vehicles would be willing to provide data for a given $f_d$. Let's model a function $v(c_1, f_d, s, V)$ as the number of vehicles that provide data to the network when the vehicles are paid $c_1$, each vehicle sends data at a frequency of $f_d$, there are $s$ servers, and a maximum of $V$ vehicles registered in the system. 

To find $v(c_1, f_d, s, V)$, we consider that the privacy sensitivity of each individual who owns a vehicle can be modeled by a random variable $e_i$, with a given cdf $F_{e_i}(\cdot)$. For an individual with sensitivity $e_i$, the utility due to privacy is given as: 

\begin{equation} \label{vutil2}
    w_i = c_1\cdot f_d - e_i\cdot L(f_d,s)
    \end{equation}

If $w_i > 0$, the individual would provide their data. The probability that this happens depends on $e_i$.  Using \cref{vutil2}, we see that for positive $w_i$ values, the probability of an individual agreeing to share their data would be:

\begin{equation} \label{vutil3}
  Pr(w_i > 0) = Pr\left(e_i < \frac{c_1 \cdot f_d}{L(f_d,s)}\right)  = F_{e_i}\left(\frac{c_1 \cdot f_d}{L(f_d,s)}\right)
    \end{equation}

For our simulations, we use a log-normal distribution for $F_{e_i}(\cdot)$. If we have $V$ total vehicles in the system, the expected number of vehicles that would participate in the network would be: 

\begin{equation} \label{vutil4}
    v(c_1, f_d, s) = V\cdot F_{e_i}\left(\frac{c_1 \cdot f_d}{L(f_d,s)}\right)
    \end{equation}
    
\vspace{5pt}

\begin{equation} \label{vutil5}
    = V\cdot\frac{1}{\sqrt{2\cdot\pi \cdot \sigma ^2}}  \cdot \frac{1}{\frac{c_1 \cdot f_d}{L(f_d,s)}} \cdot \exp\left[-\frac{\left(\ln\left(\frac{c_1 \cdot f_d}{L(f_d,s)}\right) - \mu\right)^2}{2\sigma^2}\right]
    \end{equation}

\subsection{Distributed Servers for Privacy-Preserving Data Aggregation}

We assume that each server on our network is responsible for collecting and processing the data shared by the vehicles in the network. In addition to these primary functions, the servers coordinate with each other to ensure data integrity and consistency across the system. This coordination is facilitated by a coordinator service that orchestrates aggregation rounds and task assignment among servers. In deployment, this role can be replicated across multiple nodes to avoid reliance on a single point of failure.

Each server incurs a cost proportional to the amount of data it processes, in addition to a fixed operational cost. In our proposed framework, third-party entities are responsible for server management. This keeps the data consumer separate from server administration, enhancing user privacy and trust. By distributing control, we make sure that no single entity can access the raw data of any vehicle, reinforcing our dedication to data privacy. However, the key assumption we make is that the data consumer bears the cost of the use of these servers. The total cost incurred per server would be:

\begin{equation} \label{servercost}
    c_2 \cdot v(c_1, f_d, s, V) \cdot \frac{f_d}{s} +  \frac{c_3}{s}
    \end{equation}

where $c_2$ represents the variable costs dependent on the volume of data processed, and $\frac{c_3}{s}$ allocates the coordinator’s operational costs equally among the servers.

\subsection{Economic Role of the Data Consumers}

The data consumer is the entity that provides the payments to the vehicles and servers and receives the final aggregated data. The total utility of the data consumer is, therefore, modeled as:

\begin{equation} \label{profit}
\begin{aligned}
&\quad U(v(c_1,f_d,s), f_d)\\ 
&- (c_2\cdot v(c_1,f_d,s, V)\cdot\frac{f_d}{s} + \frac{c_3}{s})\\ 
&- c_1\cdot v(c_1,f_d,s, V)\cdot f_d
\end{aligned}
\end{equation}

\Cref{profit} represents the utility gained from the vehicles' data minus the total cost to run the servers and the total payment made to the participating vehicles. 

The data consumer sets the network's parameters, deciding on the compensation for vehicles and servers, the number of servers in the network, and how often data is collected from vehicles. Based on these established conditions, vehicles can then choose to participate in the network if the terms align with their preferences.

\vspace{5pt}

\section{Economic Model} \label{econ}
\subsection{Data Consumer's Utility Function}

The data consumer's utility gained from vehicle data can be expressed as $U(v(c_1,f_d,s), f_d)$. It depends on the number of vehicles that provide data, which in turn depends on the payment made to each vehicle for providing their data $c_1$, the frequency with which these vehicles share their data $f_d$, and the total number of servers in the network $s$. 

We model the data consumer's utility in a given area as an increasing function related to the number of vehicles contributing data: 

\begin{equation} \label{util}
    -1/( 1 + a\cdot exp( \frac{-1}{\sqrt{n}}))+1
    \end{equation}

We adopt the exponential utility function in Equation~\ref{util} to align with foundational economic theory, capturing the diminishing marginal returns that occur as the volume and frequency of data contributions increase. Such diminishing returns have been extensively documented within the literature on network economics and telecommunication markets, capturing realistic consumer behavior and network utilization patterns \cite{econ_book,tele_econ}.

In \cref{util}, $n$ is the number of vehicles that contribute data and $a$ is a variable parameter that determines the limit to which the utility function converges. When $a$ increases, this convergence tends towards 1. The specific value of $a$ is less significant as long as it is sufficiently large. If $a$ becomes too small, the function will converge to a lesser value. For our demonstration, we choose $a$ to be equal to 100. 

To ensure the robustness and practical relevance of our model, we validate it using real-world data. Specifically, we employ a dataset containing timestamped coordinates from 2,927 taxis in Beijing, recorded over a 24-hour period\footnote{This dataset was obtained from the University of Southern California’s Autonomous Networks Research Group (https://anrg.usc.edu/www/downloads/)}. This dataset provides high-resolution location samples at one-minute intervals, allowing us to define the data transmission frequency $f_d$ in units of samples per minute.

\begin{figure}[!t]
    \centering
    \includegraphics[scale=0.7]{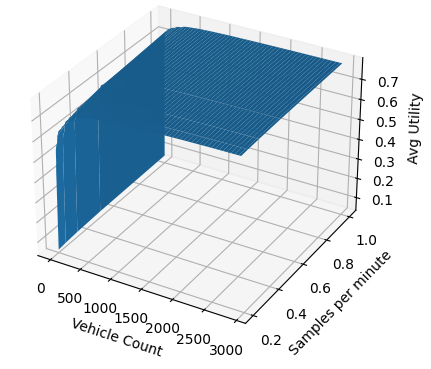}
    \caption{Data consumer's average utility as a function of the number of vehicles and the frequency of data transmission}
    \label{fig:avgutil}
\end{figure}

Our objective is to discern the data consumer's average utility if they had a subsample of this dataset. We begin by sectioning the entire dataset into spatio-temporal grids. Within each grid, we use the number of vehicle samples along with \cref{util} to calculate the utility. We then compute the average utility across all grids to determine the average utility of the dataset's region. We repeat this process with various $V$ and $f_d$ values to create a 3D plot as shown in Fig.~\ref{fig:avgutil}. This plot illustrates how the average utility for the data consumer changes as the number of vehicles and the frequency of data transmission increase. Next, we use the curve fitting function shown in \cref{uvfd} to establish an equation that expresses the utility gained from vehicle data $U(v(c_1,f_d,s), f_d)$.

\begin{equation} \label{uvfd}
    U(v(c_1,f_d,s), f_d) = \alpha \cdot(1-\exp(-\beta \cdot(v \cdot f_d)))
    \end{equation}


\vspace{5pt}

After curve fitting, our $\alpha$ and $\beta$ constants came out to be 0.99 and 0.45 respectively. We show how sensitive our results are to these parameters in Section~\ref{results}.

\subsection{Loss Function} \label{4b}

We provide a model to quantify privacy loss by evaluating how accurately a potential adversary can reconstruct a vehicle’s driving path from shared location samples. This quantification aligns with established literature on privacy economics, which emphasizes measuring privacy risks based on an adversary’s ability to infer sensitive information from partial data disclosures \cite{acquisti_economics, ghosh2011selling}. Prior studies in vehicular networks have similarly highlighted path reconstruction as a critical privacy threat, demonstrating methods to evaluate and mitigate these risks through cryptographic and SMPC techniques \cite{HEncryption, SMPC_Vanet}. Building upon these foundations, our privacy model uses path similarity as a measurable, practical proxy for privacy risk, thus allowing an economically meaningful assessment of privacy loss as a function of data sharing frequency and network structure.

Specifically, the vehicle's privacy loss function, denoted as $L(f_d,s)$, depends on the frequency with which the vehicle shares its data and the number of servers with which it shares its data. Given a set of locations on the path of a vehicle, an adversary's best estimate of the path traversed by the vehicle would be the shortest path routes between each pair of those points. Our measure of privacy quantifies how different this reconstructed path is from the vehicle's original path.

\begin{figure}[!t]
    \centering
    \includegraphics[scale = 0.32]{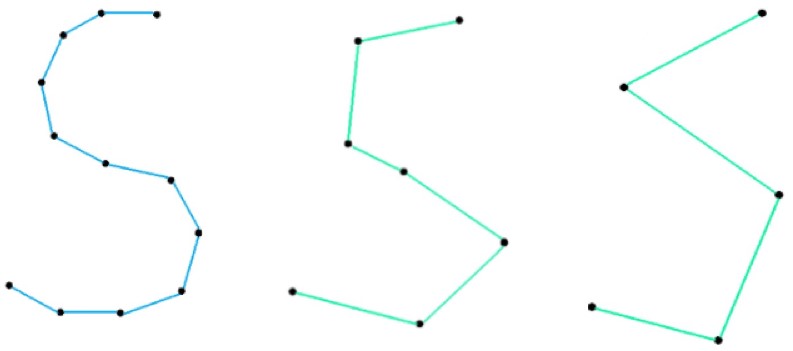}
    \caption{Paths with Different Number of Data Points}
    \label{fig:trips}
\end{figure}

Figure~\ref{fig:trips} presents an example of an original trip represented in blue, along with two subtrips, shown in green, that contain a subset of the data points from the original trip. It is evident that the paths are less similar when fewer data points are available. Consequently, paths with less similarity offer greater privacy to the user.

First, we find how much privacy is lost to one server that receives $f_d/s$ samples per minute from a given vehicle. To do this, we empirically measure our privacy metric using our dataset. We sample one trip from the dataset, which will have a $f_d$ of one sample per minute, and compare that same trip to subtrips with less frequent samples, ranging from $f_d = 1/2$ to $f_d=1/10$. We use the similarity library within the TensorBay package~\cite{tensorbay} to determine the path similarity between our initial trip and all subtrips. This library utilizes the Fr\'echet distance~\cite{frechet} as a measure of comparing two curves. Finally, we repeat this process for all vehicles in the dataset to find the average similarity scores as the value of $f_d$ changes.  

\begin{figure}[!ht]
    \centering
    \includegraphics[scale=0.25]{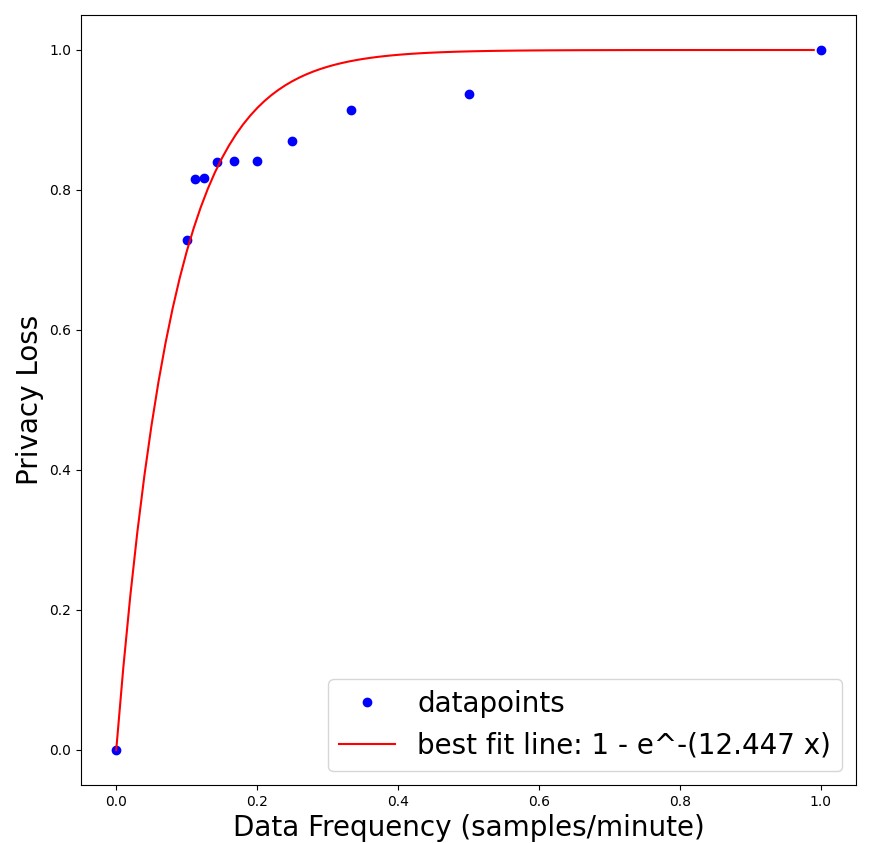}
    \caption{Privacy Loss as a function of the frequency of data transmission}
    \label{fig:similarity}
\end{figure}

Fig~\ref{fig:similarity} shows that when the frequency of data sharing increases, it becomes easier for a potential adversary to reconstruct a user's path, leading to greater privacy loss. We can also see the curve-fitted equation that gives us the privacy loss as a function of $\frac{f_d}{s}$. The loss function takes values from 0 to 1, with 1 being 100\% privacy loss. However, the network's privacy is also influenced by the total number of servers and the frequency of data transmission as well. The total loss equation is therefore:

\begin{equation}
\label{loss}
L(f_d,s) = 1 - \exp\!\left(-\frac{(r+p)f_d + q}{s}\right)
\end{equation}

\Cref{loss} models privacy loss as a bounded function taking values in $[0,1)$.
The exponential form induces monotonic increase with the data transmission frequency $f_d$ and monotonic decrease with the number of servers $s$.
The component proportional to $\frac{f_d}{s}$ captures the empirically observed degradation in privacy as the effective sampling rate per server increases, consistent with Fig.~\ref{fig:similarity}, where $r=12.447$ is obtained via least-squares curve fitting to the empirical similarity data.
The additional parameter $p$ allows flexible shaping of frequency sensitivity, while the term $\frac{q}{s}$ captures the dilution effect introduced by increasing the number of servers.
This formulation satisfies the natural limits
$\lim_{f_d\to\infty} L(f_d,s)=1 $ (for fixed $s$) and
$\lim_{s\to\infty} L(f_d,s)=0$ (for fixed $f_d$).
In the experimental evaluation, $p=0.1$ and $q=10$ are selected to appropriately balance the influence of sampling frequency and server scaling.

In addition to empirical calibration, the adopted functional form admits a natural theoretical interpretation grounded in standard inference-based privacy models. Increasing the sampling frequency $f_d$ increases the amount of observable spatiotemporal information available to a potential adversary. Under common statistical inference models, the probability of successful trajectory reconstruction grows monotonically with the number of independent observations and is frequently approximated using exponential-type convergence behavior as uncertainty is progressively reduced. The exponential structure in \Cref{loss} captures this increasing likelihood of inference as effective observation density increases.

The inverse dependence on the number of servers $s$ reflects anonymity dilution: when data is distributed across multiple aggregation points under secret-sharing, each server observes only partial information. Increasing $s$ effectively enlarges the adversarial uncertainty set, analogous to enlarging an anonymity set in $k$-anonymity–inspired privacy frameworks. Consequently, privacy loss decreases as $s$ increases.

Finally, the bounded form $L(f_d,s) \in [0,1)$ can be interpreted as a smooth surrogate for adversarial inference success likelihood, where $L \to 1$ corresponds to near-certain reconstruction and $L \to 0$ corresponds to negligible inference advantage. 

\subsection{Stackelberg Equilibrium Analysis}

In this section, we formalize the interaction between the data consumer and the vehicles as a Stackelberg game and characterize the structural properties of the resulting equilibrium.

\subsubsection*{Game Formulation}

The interaction between the data consumer and the vehicles is modeled as a two-stage Stackelberg game, where the data consumer acts as the leader and the vehicles act as followers.

In the first stage, the leader selects the strategy vector
\[
(c_1, f_d, s),
\]
where $c_1 \ge 0$ denotes the compensation per transmission unit, 
$f_d \ge 0$ denotes the data transmission frequency, and 
$s \in \mathbb{N}^+$ denotes the number of servers deployed in the network. 
For analytical purposes, we initially relax $s$ to be continuous ($s \in \mathbb{R}_+$) and later discuss the discrete case.

In the second stage, each vehicle $i$ observes the leader’s strategy and decides whether to participate based on its individual utility. 
Vehicle $i$ participates if and only if its utility in \Cref{vutil2}, evaluated at the leader’s strategy $(c_1,f_d,s)$, is non-negative.

\subsubsection*{Properties of the Participation Function}

The participation function \(v(c_1, f_d, s)\) inherits several important structural properties:

\begin{itemize}
    \item Continuity: Assuming $L(f_d,s)>0$ over the admissible design domain, and since $L(f_d,s)$ is continuous in $(f_d,s)$ and the cumulative distribution function $F_{e_i}(\cdot)$ is continuous, the composition $v(c_1,f_d,s)$ is continuous in all strategy variables.
    
    \item Monotonicity in compensation: For fixed $f_d$ and $s$, $v(c_1,f_d,s)$ is strictly increasing in $c_1$, since a higher payment relaxes the participation threshold.
    
    \item Monotonicity in server count: For fixed $c_1$ and $f_d$, $v(c_1,f_d,s)$ is increasing in $s$ because the privacy loss function $L(f_d,s)$ decreases in $s$, reducing the effective participation threshold.
\end{itemize}

These properties ensure that the induced participation response is well-defined and monotone with respect to key leader decisions.

\subsubsection*{Leader's Optimization Problem}

Given the induced participation response defined in \Cref{vutil4}, the leader solves the optimization problem stated in \Cref{profit}, anticipating the participation decision of the vehicle population.

The utility function $U(\cdot)$ defined in \Cref{uvfd} is increasing and concave in the effective data volume $v f_d$, reflecting diminishing marginal returns. The cost components are linear in $v$ and $f_d$ and decrease with larger $s$ through $1/s$ scaling.

Since the induced participation function $v(c_1,f_d,s)$ is continuous over the admissible domain (as established above), and all other components of the objective are continuous in $(c_1,f_d,s)$, the overall leader objective is continuous.

In practical deployments, compensation levels, transmission frequency, and server counts are naturally bounded due to budgetary, physical, and infrastructure constraints. Accordingly, the feasible strategy set can be restricted to a compact domain
\(
c_1 \in [0,\bar{c}],\;
f_d \in [0,\bar{f}],\;
s \in [\underline{s},\bar{s}].
\)
Over such a compact set, the continuous objective function admits a global maximizer.

\subsubsection*{Existence of Stackelberg Equilibrium}

Let $v(c_1,f_d,s)$ denote the induced participation response defined above. Substituting this response into the leader objective yields a single-level optimization problem in $(c_1,f_d,s)$, reflecting the leader’s anticipation of the followers’ participation decision.

Over a compact admissible strategy set, the resulting leader objective is continuous and therefore attains a maximizer. Let $(c_1^\star,f_d^\star,s^\star)$ denote any maximizer. Since, by construction, the participation level $v(c_1^\star,f_d^\star,s^\star)$ corresponds to vehicles optimally responding to the leader’s strategy, the pair $((c_1^\star,f_d^\star,s^\star), v(c_1^\star,f_d^\star,s^\star))$ constitutes a Stackelberg equilibrium of the two-stage game \cite{gametheory}.

Closed-form equilibrium expressions are generally not available due to the empirically calibrated privacy-loss model $L(f_d,s)$; however, equilibrium strategies can be obtained numerically by evaluating the leader's objective under the induced participation response.

\subsubsection*{Discrete Server Count}

In practice, the number of servers $s$ is discrete. 
When $s \in \mathbb{N}^+$, the leader’s problem becomes a mixed discrete–continuous optimization. 
For each fixed server level $s$, the objective function remains continuous in $(c_1, f_d)$ and admits a maximizer over a compact domain. 
Since the feasible set of $s$ is finite in practical deployments, the overall problem reduces to finitely many continuous maximization problems, and a global optimum exists by finite comparison over admissible server counts.

While discreteness may lead to piecewise variations in the objective landscape and potentially multiple local optima, it does not preclude equilibrium existence. The equilibrium can therefore be obtained by evaluating the leader’s objective across feasible server levels.

This analysis strengthens the theoretical foundation of the proposed economic framework by formally characterizing equilibrium structure while preserving the empirically grounded privacy model.

\vspace{5pt}

\section{Experiments \& Results} \label{results}
This section outlines three key experiments conducted to evaluate distinct aspects of our network's operation. The first experiment investigates how vehicle participation varies with changes in network parameters such as compensation, the number of servers, and the frequency of data sharing. The second experiment examines the network's capacity to protect user privacy by assessing how difficult it is for an adversary to reconstruct a vehicle’s path using data accessed from the network, including robustness under progressively stronger adversarial information assumptions. The third experiment focuses on measuring the network's efficiency by evaluating the latency in data retrieval, a crucial factor for supporting real-time applications.

\begin{figure*}[t]
    \centering
    \begin{minipage}[t]{0.48\textwidth}
        \includegraphics[width=\linewidth]{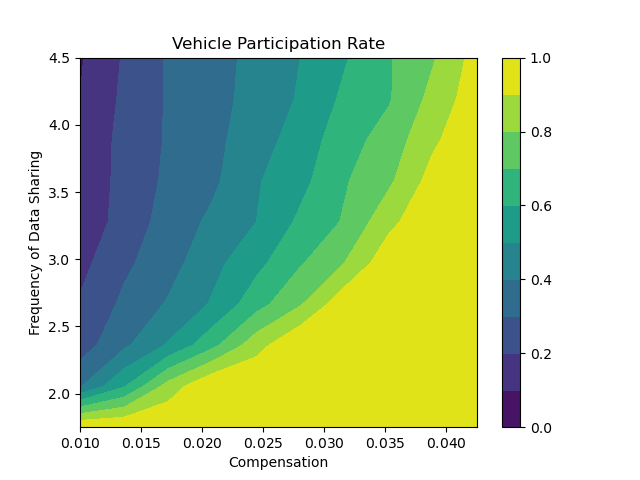}
        \caption{Vehicle participation with 5 servers in the network}
        \label{s5}
    \end{minipage}\hfill
    \begin{minipage}[t]{0.48\textwidth}
        \includegraphics[width=\linewidth]{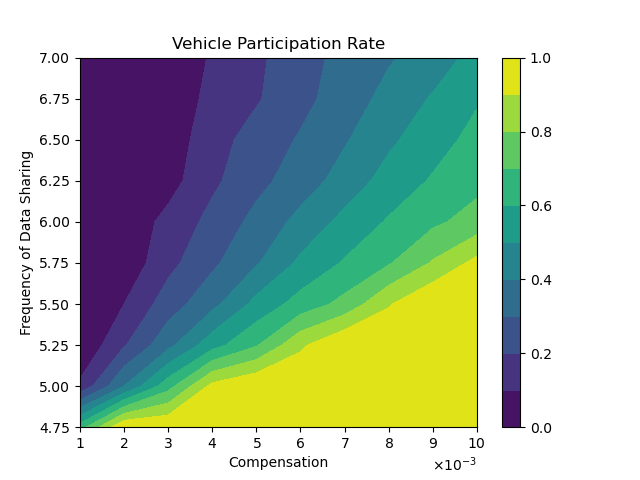}
        \caption{Vehicle participation with 10 servers in the network}
        \label{s10}
    \end{minipage}

    \begin{minipage}[t]{0.48\textwidth}
        \includegraphics[width=\linewidth]{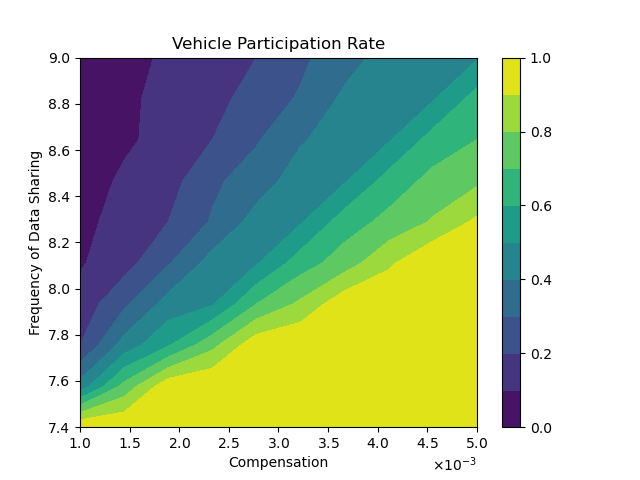}
        \caption{Vehicle participation with 15 servers in the network}
        \label{s15}
    \end{minipage}\hfill
    \begin{minipage}[t]{0.48\textwidth}
        \includegraphics[width=\linewidth]{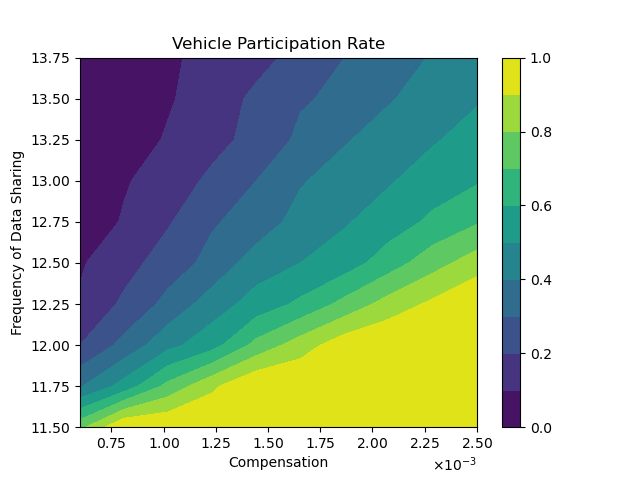} 
        \caption{Vehicle participation with 25 servers in the network}
        \label{s25}
    \end{minipage}
\end{figure*}

\subsection{Vehicle Participation}

This experiment is designed to investigate the influence of key network parameters - compensation ($c_1$), number of servers ($s$) and frequency of data sharing ($f_d$) - on vehicle participation rates. To effectively demonstrate the impact of these variables, we employ a series of four 2D contour plots, each representing a different server count configuration. These graphs are organized with compensation ($c_1$) on the x-axis and data sharing frequency ($f_d$) on the y-axis, where contour lines indicate varying levels of vehicle participation. 

Our methodology involves simulating a pool of 100 vehicles, each assigned a random privacy sensitivity level ranging from 0.1 to 0.9, with a higher value denoting a higher sensitivity to privacy concerns. A vehicle's decision to join the network is based on whether its calculated privacy score exceeds its privacy sensitivity threshold, as seen in Eq.~\ref{vutil2}. This privacy score is a crucial metric designed to reflect the perceived security and privacy protection offered by the network under specific settings of $c_1$, $f_d$, and $s$. We systematically vary $c_1$ and $f_d$ while maintaining consistent server settings for each plot, allowing us to capture and analyze the effects of these parameters on participation. Figures~\ref{s5} - \ref{s25} represent the vehicle participation rate for 5, 10, 15, and 25 servers respectively.

Our analysis of vehicle participation rates across different server configurations reveals several key insights about the system's dynamics. 

There is a clear inverse relationship between compensation ($c_1$) and the frequency of data sharing ($f_d$), where a higher $f_d$ requires an increase in $c_1$ to sustain participation. Importantly, this requirement for increased compensation grows at a slower rate with a higher server count, suggesting that robust server infrastructure can effectively mitigate the costs associated with more frequent data sharing. This observation aligns with the hypothesis that more servers enhance participation by improving data handling and security perceptions, encouraging participation even at higher $f_d$. 

Additionally, we observe that as the number of servers increases, so does vehicle participation. This trend is visually represented by the expansion of higher participation zones (lighter colored areas) in each successive plot. In particular, with more servers in the network, vehicles participate at higher frequencies of data sharing, even when compensation levels are relatively lower. This suggests that an increased server count not only accommodates more frequent data exchanges but also enhances the privacy safeguards within the network. The broader range of conditions under which vehicles are willing to share data at higher server counts indicates an improved trust in the network’s ability to protect their information, leading to higher overall participation rates. 

The data also reveals operational thresholds for server count and compensation. Although additional servers facilitate higher data sharing frequencies, the benefits in terms of participation rates plateau at a certain point, suggesting that continuous increases in server numbers are not always cost-effective. Similarly, we observe a diminishing return on increasing compensation; beyond a specific level, higher payments do not equate to proportionally higher participation rates. Recognizing these thresholds is vital for optimizing network operations, enabling precise adjustments in server deployments and compensation schemes to maximize efficiency and cost-effectiveness.

\subsection{Network Privacy}

\begin{figure*}[!t]
    \centering
    \begin{minipage}{0.49\textwidth}
        \centering
        \adjustbox{lap=-0.6cm}{ 
            \includegraphics[width=\linewidth]{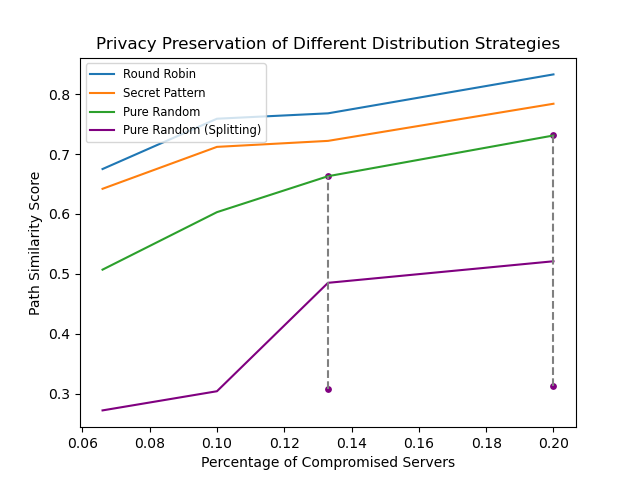}
        }
        \caption{Path similarity comparison for different data distribution strategies}
        \label{fig:strategies}
    \end{minipage}
    \hfill
    \begin{minipage}{0.49\textwidth}
        \centering
        \vspace{4mm}
        \adjustbox{lap=-1.65cm}{ 
            \includegraphics[width=\linewidth]{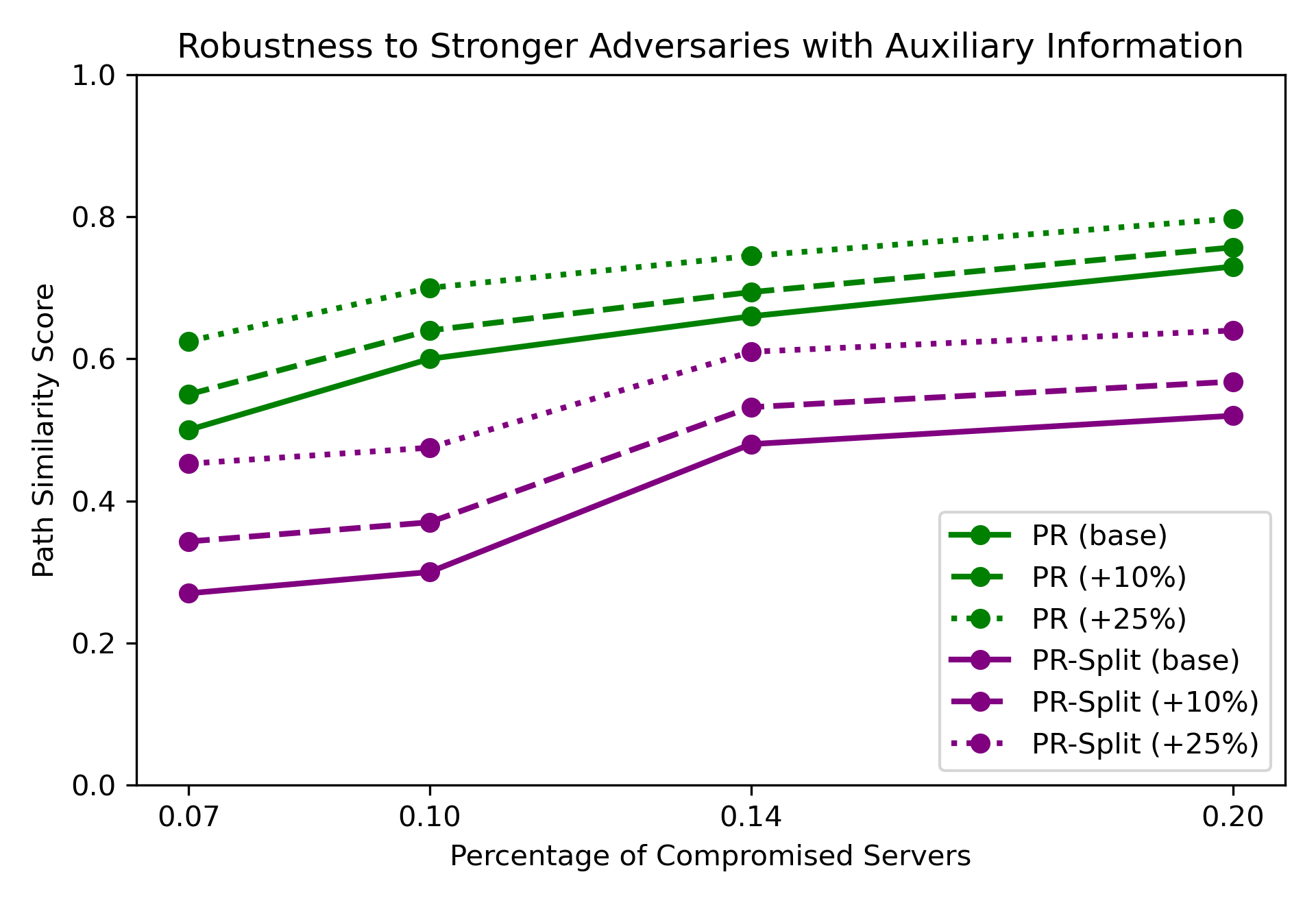}}
        \vspace{1mm}    
        \caption{Path similarity under 10\% and 25\% increased adversarial information}
        \label{fig:robustness_plot}
    \end{minipage}
\end{figure*}

This experiment aims to demonstrate the privacy-preserving capabilities of our network, specifically focusing on how well the system protects the confidentiality of vehicles' paths against potential adversarial access. As mentioned in Section~\ref{4b}, we define privacy in terms of a potential adversary's ability to accurately reconstruct a vehicle’s path if they had access to one or more servers in the network. We demonstrate how the percentage of compromised servers in the network affects the privacy of vehicle data.

We simulate a scenario where a vehicle transmits a series of data points representing its path, with each point denoted as $p_i = (x_i, y_i)$. The vehicle's data points are distributed using four different strategies: round robin, secret pattern, pure random, and pure random with splitting. For each strategy, we analyze how much of a vehicle’s path a potential adversary can reconstruct if they had access to a specific percentage of the servers in the network.

In the \textit{Round Robin} strategy, the vehicle sends its data points sequentially to a list of active servers. The first data point goes to server $s_1$, the second to server $s_2$, etc. Once the list of servers is exhausted, the process loops back to $s_1$. 

In the \textit{Secret Pattern} strategy, the vehicle and the consumer agree on a predetermined secret pattern for data distribution. For example, the $x$-coordinates might be sent in the order $(s_3, s_1, s_2)$ and the $y$-coordinates in the order $(s_2, s_3, s_1)$. For data points $(x_1, y_1), (x_2, y_2)$, the vehicle sends $x_1 \rightarrow s_3$, $y_1 \rightarrow s_2$, $x_2 \rightarrow s_1$, $y_2 \rightarrow s_3$, and so on. The consumer collects the data and rearranges them in the correct order in $O(N)$ time. 

In the \textit{Pure Random} strategy, the vehicle sends each data point to a randomly selected server, along with a timestamp. This random distribution helps obscure the pattern of the vehicle’s path. For example, data points $p_1, p_2, p_3$ might be sent to servers $s_2, s_1, s_3$ at random. The consumer needs to collect and sort the data points by timestamp in $O(N \log N)$ time to reconstruct the vehicle’s path.

In the \textit{Pure Random Splitting} strategy, the vehicle sends the $x$-coordinates and $y$-coordinates of each point to two mutually exclusive sets of servers. Each server only receives either $x$ or $y$ coordinates, but not both. For a point $(x_1, y_1)$, $x_1$ might be sent to $s_2$ and $y_1$ to $s_4$. Servers handling $x$-coordinates are different from those handling $y$-coordinates. Because of this, we assume that the adversary takes its best guess to fill in the missing coordinates. The consumer collects the data and sorts them by timestamp in $O(N \log N)$ time.

To simulate adversarial access, we model a scenario in which an adversary gains access to a defined percentage of servers within the network. We then evaluate the extent to which the adversary could reconstruct a vehicle’s original path using the compromised data. The assessment of similarity between the adversary's reconstructed path and the vehicle's actual path is conducted using the TensorBay similarity library. This tool generates a similarity score on a scale from 0 to 1, where a score of 1 indicates an exact match to the original path, representing a complete breach of privacy. In contrast, a score closer to 0 suggests effective obfuscation of the vehicle’s path, demonstrating strong privacy preservation within the network. This scoring method serves as a quantitative measure of the network's privacy integrity.


Fig.~\ref{fig:strategies} illustrates how similarity scores vary with the percentage of compromised servers across four distinct data distribution strategies. This relationship offers insights into the adversary's potential to accurately reconstruct the vehicle's route based on the compromised data.

The \textit{Round Robin} strategy exhibits the highest similarity scores, consistently performing the worst in terms of privacy protection. This outcome is likely due to the predictable nature of data dissemination in this method, which enables adversaries to reconstruct vehicle routes more easily with access to even a limited number of servers. In contrast, the \textit{Secret Pattern} strategy shows slightly better effectiveness. Although it employs a layer of unpredictability through a secret pattern, once adversaries compromise or deduce this pattern, they can reconstruct the route with greater accuracy. Compared to the first two strategies, the \textit{Pure Random} strategy provides significantly better privacy protection. The inherent randomness of this strategy makes it more difficult for adversaries to piece together accurate routes from compromised server data. 

The \textit{Pure Random Splitting} strategy emerges as the most effective strategy, consistently delivering the lowest similarity scores, thus demonstrating the strongest privacy preservation. However, this strategy exhibits variability depending on scenarios in which more than one server is compromised. If all compromised servers hold the same type of data, either exclusively $x$ or $y$ coordinates, the resulting similarity scores are lower. This outcome stems from the difficulty an adversary faces in accurately guessing the missing coordinates. In contrast, if compromised servers contain a combination of both $x$ and $y$ coordinates, the similarity scores are much higher since they are able to guess the missing coordinates with greater accuracy.

The error bars on the graph represent these two extreme scenarios: the maximum similarity scores occur when different types of data are accessible, and the minimum when only one type of coordinate is compromised. The plotted line shows the average of these extremes. The plotted line, along with the error bars, illustrates the overall variability within this strategy.


\subsubsection*{Robustness to Increased Adversarial Information}

The results presented in Fig.~\ref{fig:strategies} assume an adversary that reconstructs vehicle paths using only the data obtained from compromised servers. In realistic settings, however, adversaries may possess additional contextual information, such as common points-of-interest (POIs), commuting patterns, or temporal linkage knowledge. Such information can increase the adversary’s effective reconstruction capability and lead to higher similarity scores.

To examine the impact of stronger adversaries, we evaluate reconstruction similarity under progressively stronger adversarial information assumptions. Specifically, we consider scenarios representing moderate (10\%) and substantial (25\%) increases in available adversarial information, corresponding to increased effective inference strength. The resulting similarity scores are shown in Fig.~\ref{fig:robustness_plot} for the Pure Random and Pure Random Splitting strategies.

As expected, reconstruction similarity increases under stronger adversarial conditions. However, the qualitative behavior of the system remains consistent. Increasing the percentage of compromised servers continues to increase similarity, and the Pure Random Splitting strategy consistently maintains lower similarity scores than Pure Random across all adversarial strengths. 

These observations indicate that although stronger adversaries improve absolute reconstruction accuracy, the relative privacy advantages of server randomization and coordinate splitting persist. The proposed distribution mechanisms therefore suggest robustness to increased adversarial information.

\subsection{Network Efficiency}

To assess the efficiency and reliability of our network infrastructure, we have conducted an experiment using the Chameleon Cloud platform. This setup involves configuring multiple server instances using Docker containers, each subjected to a consistent 20ms message delay to simulate the average network conditions in a city. This experiment measures the consumer read latency in relation to both the number of servers and the volume of data samples requested.

\begin{figure}[!t]
    \centering
    \includegraphics[width=\columnwidth]{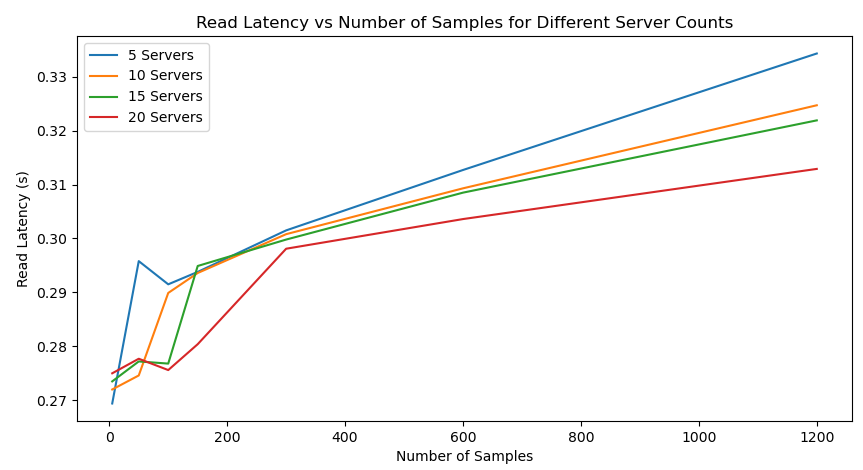}
    \caption{Consumer read latency of networks with different server counts as the number of samples increases}
    \label{fig:latency_plot}
\end{figure}

In this experiment, we have used varying server counts to observe their impact on read latency as the network's capacity scaled. The latency was recorded against an increasing number of samples to cover a broad spectrum of use cases from minimal to heavy data retrieval tasks. This approach allows for a comprehensive analysis of the network’s performance under different operational stresses. The results of this experiment are shown in Fig.~\ref{fig:latency_plot}.

Our results indicate a clear trend: higher server counts led to reduced read latencies, confirming the hypothesis that a more distributed network can handle larger loads more efficiently. The initial read latencies for a small number of samples were significantly higher across all configurations, which is attributed to the overhead involved in establishing connections and the initial data retrieval delay. However, as the number of samples increases, the latency increases more gradually, stabilizing for larger sample sizes, which illustrates the network's capability to manage sustained data transfers efficiently. These findings underscore the network's robustness in handling concurrent data requests and its scalability, crucial for applications requiring high throughput and low latency.

\section{Conclusion} \label{conclusion}
We have successfully demonstrated the practicality and effectiveness of a novel game-theoretic network framework designed to enhance privacy and scalability through decentralized data collection. By implementing distributed servers and using secure multiparty computation, our framework has proven robust in minimizing the risks associated with adversarial data reconstruction, thus significantly safeguarding vehicle data privacy.

Our economic model introduces a Stackelberg-style interaction in which the data consumer sets network parameters—including compensation, data sharing frequency, and server count—while privacy-aware vehicles decide whether to participate based on individual utility. This model enables a dynamic balance of incentives and risks.

Through simulations, we explored how these parameters interact to shape vehicle participation, network privacy, and data consumer utility. These results make explicit the trade-off at the heart of our framework: increasing data sharing frequency and payments can boost utility and participation but also increases privacy loss. Increasing the number of servers improves privacy but incurs additional infrastructure costs. Our experiments empirically illustrate how data consumers can navigate this trade-off to balance profit and privacy—a core goal of the system.

Unlike many game-theoretic models that solve for closed-form equilibria, our approach is simulation-driven. Rather than deriving optimal strategies analytically, we provide a flexible framework that enables stakeholders to explore different configurations and outcomes under realistic privacy preferences. This makes the framework adaptable to real-world deployments, especially when privacy loss functions are nonlinear or empirically defined.

Future studies can explore transitioning from our current hybrid model to a fully decentralized architecture, which could significantly improve security and privacy. Furthermore, applying this framework to blockchain technology and testing its authenticity and reliability via smart contracts represent important directions for advancement. This strategy would confirm the framework's effectiveness in real-world settings and leverage blockchain to foster transparency and trust among stakeholders.

In conclusion, this paper bridges the technical and economic dimensions of data sharing in vehicular networks. It contributes a simulation-based, privacy-aware platform grounded in game-theoretic principles, offering both practical relevance and theoretical depth for future research in secure, incentivized data markets.

\section*{Acknowledgments}

This manuscript was edited for grammatical accuracy and clarity using artificial intelligence tools, which did not contribute to any original content creation.

\bibliographystyle{IEEEtran}
\bibliography{ref}

@article{Ogonji,
    title = {A survey on privacy and security of {Internet} of {Things}},
    journal = {Computer Science Review},
    volume = {38},
    issn = {1574-0137},
    doi = {10.1016/j.cosrev.2020.100312},    author = {Ogonji, Mark and Okeyo, George and Wafula, Joseph Muliaro},
    month = nov,
    year = {2020},    pages = {100312}
}

@article{HEncryption,
    title = {Use of {Homomorphic} {Encryption} with {GPS} in {Location} {Privacy}},
    doi = {10.1109/ISCON47742.2019.9036149},
    journal = {2019 4th International Conference on Information Systems and Computer Networks (ISCON)},
    author = {Gupta, Shreya and Arora, Ginni},
    month = nov,
    year = {2019},
    pages = {42--45}
}

@article{SMPC,
    title = {Secure {Multi}-{Party} {Computation}: {Theory}, practice and applications},
    journal = {Information Sciences},
    volume = {476},
    issn = {0020-0255},
    shorttitle = {Secure {Multi}-{Party} {Computation}},
    doi = {10.1016/j.ins.2018.10.024},    author = {Zhao, Chuan and Zhao, Shengnan and Zhao, Minghao and Chen, Zhenxiang and Gao, Chong-Zhi and Li, Hongwei and Tan, Yu-an},
    month = feb,
    year = {2019},    pages = {357--372}
}

@article{SMPC_Vanet,
author = {Song, Chuanyuan and Zhang, M.Y. and Peng, W.P.},
year = {2018},
month = {01},
pages = {99-107},
title = {Research on secure and privacy-preserving scheme based on secure multi-party computation for VANET},
volume = {9},
journal = {Journal of Information Hiding and Multimedia Signal Processing}
}

@inproceedings{AI_Blockchain_IoV,
    title = {{AI}-{Powered} {Blockchain} - {A} {Decentralized} {Secure} {Multiparty} {Computation} {Protocol} for {IoV}},
    doi = {10.1109/INFOCOMWKSHPS50562.2020.9162866},
    booktitle = {{IEEE} {Conference} on {Computer} {Communications} {Workshops}},
    author = {Raja, Gunasekaran and Manaswini, Yelisetty and Vivekanandan, Gaayathri Devi and Sampath, Harish and Dev, Kapal and Bashir, Ali Kashif},
    month = jul,
    year = {2020},    pages = {865--870}
}

@inproceedings{AutoMPC,
author = {Li, Tao and Lin, Lei and Gong, Siyuan},
year = {2019},
month = {01},
pages = {},
booktitle={SafeAI@AAAI},
title = {AutoMPC: Efficient Multi-Party Computation for Secure and Privacy-Preserving Cooperative Control of Connected Autonomous Vehicles}
}

@article{blockchain_meets_vanet,
    title = {Blockchain {Meets} {VANET}: {An} {Architecture} for {Identity} and {Location} {Privacy} {Protection} in {VANET}},
    volume = {12},
    issn = {1936-6442, 1936-6450},
    shorttitle = {Blockchain {Meets} {VANET}},
    doi = {10.1007/s12083-019-00786-4},    number = {5},
    journal = {Peer-to-Peer Networking and Applications},
    author = {Li, Hui and Pei, Lishuang and Liao, Dan and Sun, Gang and Xu, Du},
    month = sep,
    year = {2019},
    pages = {1178--1193}
}

@article{Blockchain_solutions_IoV,
    title = {Privacy-{Preserving} {Solutions} in {Blockchain}-{Enabled} {Internet} of {Vehicles}},
    volume = {11},
    issn = {2076-3417},
doi = {10.3390/app11219792},    journal = {Applied Sciences},
    author = {Kaltakis, Konstantinos and Polyzi, Panagiota and Drosatos, George and Rantos, Konstantinos},
    month = oct,
    year = {2021},
    pages = {9792}
}

@article{bc_iov_survey,
    title = {Survey on blockchain-based applications in internet of vehicles}, 
    journal = {Computer and Electrical Engineering},
    volume = {84},
    issn = {0045-7906},
    doi = {10.1016/j.compeleceng.2020.106646},    
    author = {Mendiboure, Leo and Chalouf, Mohamed Aymen and Krief, Francine},
    month = jun,
    year = {2020},    pages = {106646}
}

@article{jiang_blockchain-based_2019,
    title = {Blockchain-{Based} {Internet} of {Vehicles}: {Distributed} {Network} {Architecture} and {Performance} {Analysis}},
    volume = {6},
    issn = {2327-4662},
    shorttitle = {Blockchain-{Based} {Internet} of {Vehicles}},
    doi = {10.1109/JIOT.2018.2874398},    number = {3},
    journal = {IEEE Internet of Things Journal},
    author = {Jiang, Tigang and Fang, Hua and Wang, Honggang},
    month = jun,
    year = {2019},    pages = {4640--4649}
}

@INPROCEEDINGS{security_challenges_iov,
  author={Dai Nguyen, Huu Phuoc and Zoltán, Rajnai},
  booktitle={2018 IEEE 16th International Symposium on Intelligent Systems and Informatics (SISY)}, 
  title={The Current Security Challenges of Vehicle Communication in the Future Transportation System}, 
  year={2018},
  doi={10.1109/SISY.2018.8524773}}

@article{blockchain_security_services,
    title = {Security {Services} {Using} {Blockchains}: {A} {State} of the {Art} {Survey}},
    volume = {21},
    issn = {1553-877X},
    shorttitle = {Security {Services} {Using} {Blockchains}},
    doi = {10.1109/COMST.2018.2863956},    number = {1},
    journal = {IEEE Communications Surveys \& Tutorials},
    author = {Salman, Tara and Zolanvari, Maede and Erbad, Aiman and Jain, Raj and Samaka, Mohammed},
    year = {2019},    pages = {858--880}
}

@inproceedings{smartcontracts,
    title = {An {Overview} of {Smart} {Contract}: {Architecture}, {Applications}, and {Future} {Trends}},
    shorttitle = {An {Overview} of {Smart} {Contract}},
    doi = {10.1109/IVS.2018.8500488},    booktitle = {2018 {IEEE} {Intelligent} {Vehicles} {Symposium} ({IV})},
    author = {Wang, Shuai and Yuan, Yong and Wang, Xiao and Li, Juanjuan and Qin, Rui and Wang, Fei-Yue},
    month = jun,
    year = {2018},    pages = {108--113}
}

@article{IoV,
author = {Lee, Eun-Kyu and Gerla, Mario and Pau, Giovanni and Lee, Uichin and Lim, Jae-Han},
year = {2016},
month = {09},
pages = {},
title = {Internet of Vehicles: From intelligent grid to autonomous cars and vehicular fogs},
volume = {12},
journal = {International Journal of Distributed Sensor Networks},
doi = {10.1177/1550147716665500}
}

@ARTICLE{Hao,  author={Ye, Hao and Liang, Le and Ye Li, Geoffrey and Kim, JoonBeom and Lu, Lu and Wu, May},  journal={IEEE Vehicular Technology Magazine},   title={Machine Learning for Vehicular Networks: Recent Advances and Application Examples},   year={2018},  volume={13},  number={2},  pages={94-101},  doi={10.1109/MVT.2018.2811185}}

@INPROCEEDINGS{PoL,
  author={Amoretti, Michele and Brambilla, Giacomo and Medioli, Francesco and Zanichelli, Francesco},
  booktitle={2018 IEEE International Conference on Software Quality, Reliability and Security Companion (QRS-C)}, 
  title={Blockchain-Based Proof of Location}, 
  year={2018},
  volume={},
  number={},
  pages={146-153},
  doi={10.1109/QRS-C.2018.00038}}

@misc{alladi2022comprehensive,
      title={A Comprehensive Survey on the Applications of Blockchain for Securing Vehicular Networks}, 
      author={Tejasvi Alladi and Vinay Chamola and Nishad Sahu and Vishnu Venkatesh and Adit Goyal and Mohsen Guizani},
      year={2022},
      eprint={2201.04803},
      archivePrefix={arXiv},
      primaryClass={cs.CR}
}

@InProceedings{Freedman,
author="Freedman, Michael J.
and Nissim, Kobbi
and Pinkas, Benny",
editor="Cachin, Christian
and Camenisch, Jan L.",
title="Efficient Private Matching and Set Intersection",
booktitle="Advances in Cryptology - EUROCRYPT 2004",
year="2004",
publisher="Springer Berlin Heidelberg",
address="Berlin, Heidelberg",
pages="1--19",
isbn="978-3-540-24676-3"
}

@article{genome,
author={Ma,X. and Li,J. and Zhang,F.},
year={2015},
title={Refereed computation delegation of private sequence comparison in cloud computing},
journal={International Journal of Network Security},
volume={17},
number={6},
pages={743-753},
}

@article{bost2014machine,
  title={Machine learning classification over encrypted data},
  author={Bost, Raphael and Popa, Raluca Ada and Tu, Stephen and Goldwasser, Shafi},
  journal={Cryptology ePrint Archive},
  year={2014}
}

@book{homomorphic,
  title={Homomorphic encryption},
  author={Yi, Xun and Paulet, Russell and Bertino, Elisa and Yi, Xun and Paulet, Russell and Bertino, Elisa},
  year={2014},
  publisher={Springer}
}

@article{bc_datasharing,
  title={A Secure and Efficient Blockchain-Based Data Sharing Scheme for Location Data},
  author={Zirui Hu and Yuhan Yang and Jing Wu and Chengnian Long},
  journal={The 2022 4th International Conference on Blockchain Technology},
  year={2022}
}

@article{Javed2020BlockchainBasedSD,
  title={Blockchain-Based Secure Data Storage for Distributed Vehicular Networks},
  author={Muhammad Umar Javed and Mubariz Rehman and Nadeem Javaid and Abdulaziz Aldegheishem and Nabil Ali Alrajeh and Muhammad Tahir},
  journal={Applied Sciences},
  year={2020}
}

@article{frechet,
  title={Computing the Fr{\'e}chet distance between two polygonal curves},
  author={Alt, Helmut and Godau, Michael},
  journal={International Journal of Computational Geometry \& Applications},
  volume={5},
  pages={75--91},
  year={1995},
  publisher={World Scientific},
doi = {10.1142/S0218195995000064}
}

@article{chen2022,
  title={A Summary of Security Techniques-Based Blockchain in IoV},
  author={Chen, Chen and Quan, Shi},
  journal={Security and Communication Networks},
  volume={2022},
  year={2022},
  publisher={Hindawi}
}

@article{garg2020survey,
  title={A survey on security and privacy issues in IoV.},
  author={Garg, Tanvi and Kagalwalla, Navid and Churi, Prathamesh and Pawar, Ambika and Deshmukh, Sanjay},
  journal={International Journal of Electrical \& Computer Engineering},
  volume={10},
  number={5},
  year={2020}
}

@book{gametheory,
  title={An introduction to game theory},
  author={Osborne, Martin J and others},
  volume={3},
  year={2004},
  publisher={Oxford university press New York}
}

@article{gupta_game_2022,
    title = {Game {Theory}-{Based} {Authentication} {Framework} to {Secure} {Internet} of {Vehicles} with {Blockchain}},
    volume = {22},
    copyright = {http://creativecommons.org/licenses/by/3.0/},
    issn = {1424-8220},
    doi = {10.3390/s22145119},    number = {14},
    journal = {Sensors},
    author = {Gupta, Manik and Kumar, Rakesh and Shekhar, Shashi and Sharma, Bhisham and Patel, Ram Bahadur and Jain, Shaily and Dhaou, Imed Ben and Iwendi, Celestine},
    month = jan,
    year = {2022},
    pages = {5119}
}

@article{liwang_game_2019,
    title = {Game {Theory} {Based} {Opportunistic} {Computation} {Offloading} in {Cloud}-{Enabled} {IoV}},
    volume = {7},
    issn = {2169-3536},
    doi = {10.1109/ACCESS.2019.2897617},    journal = {IEEE Access},
    author = {Liwang, Minghui and Wang, Jiexiang and Gao, Zhibin and Du, Xiaojiang and Guizani, Mohsen},
    year = {2019},    pages = {32551--32561}
}

@ARTICLE{Xu,
  author={Xu, Xiaolong and Jiang, Qinting and Zhang, Peiming and Cao, Xuefei and Khosravi, Mohammad R. and Alex, Linss T. and Qi, Lianyong and Dou, Wanchun},
  journal={IEEE Transactions on Fuzzy Systems}, 
  title={Game Theory for Distributed IoV Task Offloading With Fuzzy Neural Network in Edge Computing}, 
  year={2022},
  volume={30},
  number={11},
  pages={4593-4604},
  doi={10.1109/TFUZZ.2022.3158000}}

@article{hassija_dagiov_2020,
    title = {{DAGIoV}: {A} {Framework} for {Vehicle} to {Vehicle} {Communication} {Using} {Directed} {Acyclic} {Graph} and {Game} {Theory}},
    volume = {69},
    issn = {1939-9359},
    shorttitle = {{DAGIoV}},
    doi = {10.1109/TVT.2020.2968494},    number = {4},
    journal = {IEEE Transactions on Vehicular Technology},
    author = {Hassija, Vikas and Chamola, Vinay and Han, Guangjie and Rodrigues, Joel J. P. C. and Guizani, Mohsen},
    month = apr,
    year = {2020},    pages = {4182--4191}
}

@ARTICLE{Chen,
  author={Chen, Xu and Jiao, Lei and Li, Wenzhong and Fu, Xiaoming},
  journal={IEEE/ACM Transactions on Networking}, 
  title={Efficient Multi-User Computation Offloading for Mobile-Edge Cloud Computing}, 
  year={2016},
  volume={24},
  number={5},
  pages={2795-2808},
  doi={10.1109/TNET.2015.2487344}}

@ARTICLE{He,
  author={He, Qiang and Cui, Guangming and Zhang, Xuyun and Chen, Feifei and Deng, Shuiguang and Jin, Hai and Li, Yanhui and Yang, Yun},
  journal={IEEE Transactions on Parallel and Distributed Systems}, 
  title={A Game-Theoretical Approach for User Allocation in Edge Computing Environment}, 
  year={2020},
  volume={31},
  number={3},
  pages={515-529},
  doi={10.1109/TPDS.2019.2938944}}

@article{lu_blockchain_2020,
    title = {Blockchain {Empowered} {Asynchronous} {Federated} {Learning} for {Secure} {Data} {Sharing} in {Internet} of {Vehicles}},
    volume = {69},
    issn = {1939-9359},
    doi = {10.1109/TVT.2020.2973651},    number = {4},
    journal = {IEEE Transactions on Vehicular Technology},
    author = {Lu, Yunlong and Huang, Xiaohong and Zhang, Ke and Maharjan, Sabita and Zhang, Yan},
    month = apr,
    year = {2020},    pages = {4298--4311}
}

@article{fan_cloud-based_2021,
    title = {Cloud-based {RFID} mutual authentication scheme for efficient privacy preserving in {IoV}},
    volume = {358},
    issn = {0016-0032},
    doi = {10.1016/j.jfranklin.2019.02.023},    number = {1},
    journal = {Journal of the Franklin Institute},
    author = {Fan, Kai and Jiang, Wei and Luo, Qi and Li, Hui and Yang, Yintang},
    month = jan,
    year = {2021},
    pages = {193--209}
}

@article{kong_privacy-preserving_2019,
    title = {A privacy-preserving sensory data sharing scheme in {Internet} of {Vehicles}},
    volume = {92},
    issn = {0167-739X},
    doi = {10.1016/j.future.2017.12.003},    journal = {Future Generation Computer Systems},
    author = {Kong, Qinglei and Lu, Rongxing and Ma, Maode and Bao, Haiyong},
    month = mar,
    year = {2019},
    pages = {644--655}
}

@article{Kaiser,
    title = {Towards a {Privacy}-{Preserving} {Way} of {Vehicle} {Data} {Sharing} – {A} {Case} for {Blockchain} {Technology}?},
    doi = {10.1007/978-3-319-99762-9_10},
    author = {Kaiser, Christian and Steger, Marco and Dorri, Ali and Festl, Andreas and Stocker, Alexander and Fellmann, Michael and Kanhere, Salil},
    journal = {Advanced Microsystems for Automotive Applications 2018},
    editor = {Dubbert, Jörg and Müller, Beate and Meyer, Gereon},
    year = {2019},
    doi = {10.1007/978-3-319-99762-9_10},
    pages = {111--122}
}

@article{zhang_blockchain-based_2022,
    title = {Blockchain-based anonymous authentication for traffic reporting in {VANETs}},
    volume = {34},
    issn = {0954-0091},
    doi = {10.1080/09540091.2022.2026888},
    number = {1},
    journal = {Connection Science},
    author = {Zhang, Li and Xu, Jianbo},
    month = dec,
    year = {2022},
    pages = {1038--1065}
}

@article{yeh_blockchain-based_2022,
    title = {Blockchain-{Based} {Privacy}-{Preserving} and {Sustainable} {Data} {Query} {Service} {Over} {5G}-{VANETs}},
    volume = {23},
    issn = {1558-0016},
    doi = {10.1109/TITS.2022.3146322},    number = {9},
    journal = {IEEE Transactions on Intelligent Transportation Systems},
    author = {Yeh, Lo-Yao and Shen, Nong-Xiang and Hwang, Ren-Hung},
    month = sep,
    year = {2022},    pages = {15909--15921}
}

@article{ameen2020review,
  title={A review on vehicle to vehicle communication system applications},
  author={Ameen, Hussein Ali and Mahamad, AK and Saon, S and Nor, D Md and Ghazi, K},
  journal={Indonesian Journal of Electrical Engineering and Computer Science},
  volume={18},
  number={1},
  pages={188--198},
  year={2020}
}

@article{Soto2021ASO,
  title={A survey on road safety and traffic efficiency vehicular applications based on C-V2X technologies},
  author={Ignacio Soto and Mar{\'i}a Calder{\'o}n and Oscar Amador and Manuel Urue{\~n}a},
  journal={Vehicular Communications},
  year={2021}
}

@ARTICLE{cellular_v2x,
  author={Abou-zeid, Hatem and Pervez, Farhan and Adinoyi, Abdulkareem and Aljlayl, Mohammed and Yanikomeroglu, Halim},
  journal={IEEE Consumer Electronics Magazine}, 
  title={Cellular V2X Transmission for Connected and Autonomous Vehicles Standardization, Applications, and Enabling Technologies}, 
  year={2019},
  volume={8},
  number={6},
  pages={91-98},
  doi={10.1109/MCE.2019.2941467}}

@misc{tensorbay,
  author       = {{Graviti}},
  title        = {TensorBay Documentation},
  year         = {2021},
  howpublished = {\url{https://tensorbay-python-sdk.graviti.com/en/v1.24.2/reference/api/geometry/polyline.html}}
}

@INPROCEEDINGS{alsaqabi,
  author={AlSaqabi, Yousef and Krishnamachari, Bhaskar},
  booktitle={2023 IEEE 98th Vehicular Technology Conference (VTC2023-Fall)}, 
  title={Incentivizing Private Data Sharing in Vehicular Networks: A Game-Theoretic Approach}, 
  year={2023},
  volume={},
  number={},
  pages={1-8},  doi={10.1109/VTC2023-Fall60731.2023.10333865}}

@article{alsaqabi-hci,
      title={Driving with Guidance: Exploring the Trade-Off Between GPS Utility and Privacy Concerns Among Drivers}, 
      author={Yousef AlSaqabi and Souti Chattopadhyay},
      journal = {arXiv:2309.12601},
      year={2023}
}

@inproceedings{LSS,
  title={General Secure Multi-party Computation from any Linear Secret-Sharing Scheme},
  author={Ronald Cramer and Ivan Damg{\aa}rd and Ueli Maurer},
  booktitle={International Conference on the Theory and Application of Cryptographic Techniques},
  year={2000}
}

@inproceedings{smpc_blockchain,
  title={Secure multi-party computation on blockchain: An overview},
  author={Zhong, Hanrui and Sang, Yingpeng and Zhang, Yongchun and Xi, Zhicheng},
  booktitle={Parallel Architectures, Algorithms and Programming: 10th International Symposium, PAAP 2019, Guangzhou, China, December 12--14, 2019, Revised Selected Papers 10},
  pages={452--460},
  year={2020},
  organization={Springer}
}

@inproceedings{block_smpc,
  title={Block-smpc: A blockchain-based secure multi-party computation for privacy-protected data sharing},
  author={Yang, Yuhan and Wei, Lijun and Wu, Jing and Long, Chengnian},
  booktitle={Proceedings of the 2020 2nd International Conference on Blockchain Technology},
  pages={46--51},
  year={2020}
}

@article{game_theory_vanet,
  author       = {Zemin Sun and
                  Yanheng Liu and
                  Jian Wang and
                  Anil Carie and
                  Dongpu Cao},
  title        = {Game Theoretic Approaches in Vehicular Networks: {A} Survey},
  journal      = {CoRR},
  volume       = {abs/2006.00992},
  year         = {2020},  eprinttype    = {arXiv},
  eprint       = {2006.00992},
  bibsource    = {dblp computer science bibliography, https://dblp.org}
}

@article{tele_econ,
author = {Shakkottai, Srinivas and Srikant, R.},
title = {Economics of network pricing with multiple ISPs},
year = {2006},
issue_date = {December 2006},
publisher = {IEEE Press},
volume = {14},
number = {6},
issn = {1063-6692},doi = {10.1109/TNET.2006.886393},
journal = {IEEE/ACM Trans. Netw.},
month = dec,
pages = {1233–1245},
numpages = {13}
}

@book{econ_book,
  title={Economics},
  author={Samuelson, Paul and Nordhaus, William},
  year={2009},
  publisher={McGraw Hill}
}

@article{acquisti_economics,
author = {Acquisti, Alessandro and Taylor, Curtis and Wagman, Liad},
year = {2016},
month = {06},
pages = {442-492},
title = {The Economics of Privacy},
volume = {54},
journal = {Journal of Economic Literature},
doi = {10.1257/jel.54.2.442}
}

@inproceedings{ghosh2011selling,
  title={Selling privacy at auction},
  author={Ghosh, Arpita and Roth, Aaron},
  booktitle={Proceedings of the 12th ACM conference on Electronic commerce},
  pages={199--208},
  year={2011}
}


\begin{IEEEbiography}[{\includegraphics[width=1in,height=1.25in,clip,keepaspectratio]{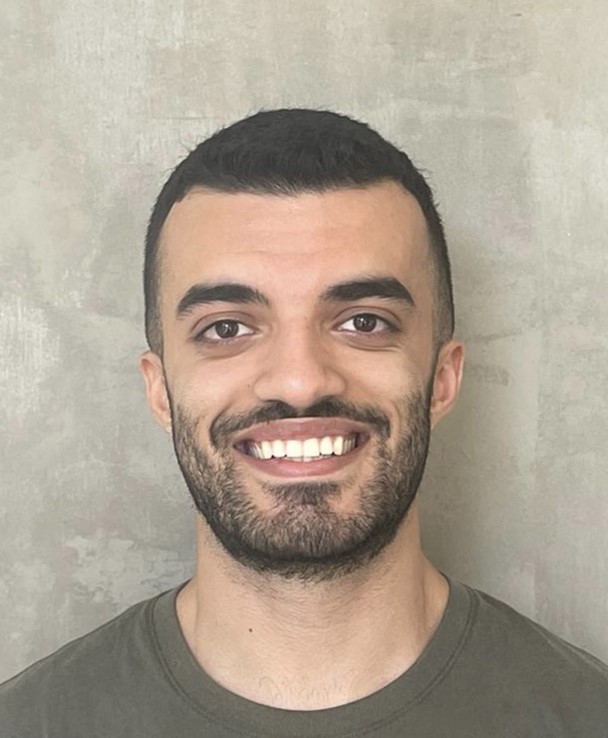}}]{Yousef AlSaqabi} is an Assistant Professor at Kuwait University. He received his PhD in Electrical Engineering from the University of Southern California in 2024. He received both his BS and MS degrees in Electrical Engineering from The Pennsylvania State University in 2018. His research focuses on the privacy and security of vehicular networks and the application of reinforcement learning to optimize route suggestions for autonomous vehicles. 

\end{IEEEbiography}

\begin{IEEEbiography}[{\includegraphics[width=1in,height=1.25in,clip,keepaspectratio]{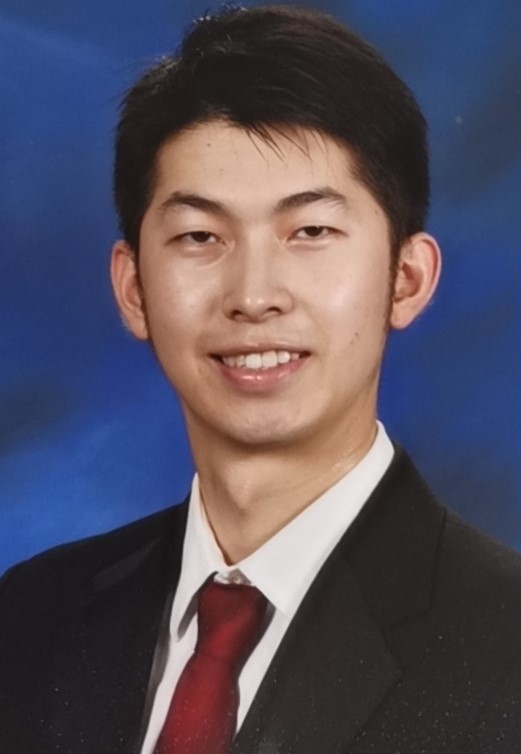}}]{Yinan Zhou}
is currently pursuing a PhD in Computer Science at the University of California, Irvine. He received his BS and MS degrees in Computer Science from the University of California, Irvine, in 2020 and 2022. His research interests include blockchains, decentralized databases, zero-knowledge proof, IoT, and edge-cloud computation networks.
\end{IEEEbiography}

\begin{IEEEbiography}[{\includegraphics[width=1in,height=1.25in,clip,keepaspectratio]{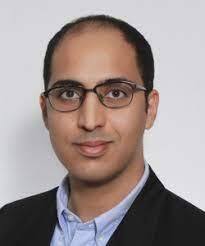}}]{Faisal Nawab}
is an Assistant Professor at the University of California, Irvine (UCI). He leads EdgeLab, which tackles research problems in the intersection of data management and distributed systems with a focus on decentralized and Internet of Things (IoT) applications. He has published papers in VLDB, SIGMOD, ICDE, EDBT, IEEE IoT, and other data management and systems conferences
and journals.
\end{IEEEbiography}

\begin{IEEEbiography}[{\includegraphics[width=1in,height=1.25in,clip,keepaspectratio]{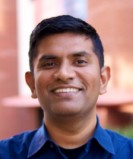}}]{Bhaskar Krishnamachari}
received the BE degree in electrical engineering from the Cooper Union in New York City in 1998 and the MS and PhD degrees in electrical engineering from Cornell University in 1999 and 2002, respectively. He is the Ming Hsieh Faculty fellow and professor of electrical engineering and computer science at the Viterbi School of Engineering at the University of Southern California. He is the Director of the Autonomous Networks Research Group and Co-Director of the Ming Hsieh Institute. He works on the design and analysis of algorithms, protocols, and applications for the Internet of Things and other distributed systems.
\end{IEEEbiography}

\end{document}